# A NOVEL PROCESS MAPPING STRATEGY IN CLUSTERED ENVIRONMENTS

Mohsen Soryani and Morteza Analoui and Ghobad Zarrinchian

Department of Computer Engineering, Iran University of Science and Technology, Tehran, Iran
`{soryani,analoui,zarrinchian}@iust.ac.ir`

## ABSTRACT

*Nowadays the number of available processing cores within computing nodes which are used in recent clustered environments, are growing up with a rapid rate. Despite this trend, the number of available network interfaces in such computing nodes has almost been remained unchanged. This issue can lead to high usage of network interface in many workloads, especially in heavy-communicating workloads. As a result, network interface may raise as a performance bottleneck and can drastically degrade the performance. The goal of this paper is to introduce a new process mapping strategy in multi-core clusters aimed at reducing network interface contention and improving inter-node communication performance of parallel applications. Performance evaluation of the new mapping algorithm in synthetic and real workloads indicates that the new strategy can achieve 5% to 90% performance improvement in heavy communicating workloads, compared to other well-known methods.*

## KEYWORDS

*Multi-Core Clusters, Network Interface, Process Mapping, Inter-Node Communication*

## 1. INTRODUCTION

Parallel processing and the use of multi-core processor chips have been major milestones in computing era since reaching uni-core chips to their performance limitations. The need to have more computational capabilities for business and scientific applications has dictated more and more processing power and this trend has never stopped. Currently, multi-core processors have experienced many advancements and this industry is known as a mature field. Although multi-core and many-core architectures have been used as main architectures for some small-scale applications but these architectures are currently used as building blocks of wider parallel architectures. Cluster computing and grid computing are some examples of such wider architectures which are well-known distributed systems. Among current distributed systems, clustered environments have gained much popularity in the field of high performance processing (HPC) such that based on [1] more than 82% of 500 top supercomputers in the world use this kind of distributed systems. The combination of cluster computing systems together with multi-core architectures is a promising way to obtain extremely high performance when running grand challenge applications such as weather forecasting and molecular dynamic modeling.

Although multi-core processors can improve computational capabilities but they also introduce some important challenges. The main challenge in this regard, is the contention of various physical cores for using shared resources like memory and system's buses. With the presence of such contention, shared resources may raise as performance bottlenecks and hence substantially degrade the performance of parallel applications. Consequently, efficient execution of parallel applications in such environments needs more deliberations of them. In doing so, there are a lot of studies including [2-6] which provide insights into the conditions in which efficient performance of multi-core clustered systems can be gained.

When multi-core computing nodes are used singularly, memory units and system's buses are the main shared resources that contending for them can adversely affect the performance. But when multi-core architectures are used as building blocks of clustered environments in which many computing nodes are connected together, network interfaces are raised as another important shared resource. This is because various processes of a parallel job (when placed on different computing nodes) use these interfaces for their communications and synchronizations. In spite of considerable growth in the number of processing cores within recent computing nodes, the number of available network interfaces has almost been remained unchanged and this number is 1 or 2 for most systems. This issue can lead to high usage of network interfaces in many workloads, especially in workloads which have high communication volume among their processes. Since a network interface can service just one request at a time, other communication requests received from different physical cores must be queued to service later. The more cores in a node, the more requests for the network interface. As a result, waiting time of messages at interface queue will be increased. This issue can finally prolong the execution time of parallel applications. Based on these issues, if we distribute parallel processes in available computing nodes such that requests arriving to each network interface be decreased, queuing time of messages at interface queues will be decreased as well and we can expect improvement in performance. Pursuing this purpose, our goal in this paper, is to present a solution for mapping parallel processes to multi-core clusters so as to reduce network interface contention. For performance evaluation of proposed mapping strategy we define some synthetic and real workloads and we show that the new approach can obtain 5% to 90% performance improvement in simulated scenarios, compared to some other well-known methods.

## 2. PAPER ORGANIZATION

This paper is organized as follows: In section 3 some related works regarding the subject is presented. In section 4 we present our solution to reduce network interface contention in multi-core cluster environments. The performance results of the proposed solution together with their comparisons with other methods in synthetic and real workloads are presented in section 5. A conclusion to this study is explained in section 6 and finally, some aspects of extending the work is presented as future works in section 7.

## 3. RELATED WORKS

There are many studies which deal with improving communication performance in clustered environments. These studies propose different approaches to reach the goal. Some works focus on improving MPI libraries for enhancing the communications. An example of such works is [7]. Some other studies offer communication improvement for particular interconnection networks. An example of such studies is [8] which tries to improve communication performance in InfiniBand interconnection networks. In this study, the data to be sent is placed in large pages so that less page address translation is required. As a result, the required time for memory registration operations and hence, communication latency is reduced. But the most effective approach for enhancing communication performance is to use efficient process mapping techniques. This approach has attracted more attentions compared to two other mentioned methods.

Various methods have been proposed for mapping parallel processes to processing elements. Among these methods, Blocked and Cyclic are two common approaches which are already investigated in [9-10]. In Blocked technique, the mapping procedure is started by selecting a computing node and assigning parallel processes to its free cores one-by-one. When there is no free core in the selected node, another computing node is selected and this procedure is repeated until the end of assignment. In the Cyclic method, parallel processes are distributed among computing nodes in a Round Robin fashion. As a result, maximum number of nodes and minimum number of cores in each node is used in this method (in contrast to the Blocked which uses minimum number of nodes and maximum number of cores in each node).

Although Blocked and Cyclic methods are practically the default method in many situations, but these approaches have little intelligence and do not consider the volume of communications between parallel processes. Because of this issue, other techniques have been proposed which are more intelligent than the Blocked and Cyclic. Some of these methods are [11-15]. Proposed mapping strategy in these studies is based on graph partitioning techniques. The main idea in these techniques is to find processes which communicate to each other frequently and to map them near each other as much as possible (e.g. to place them in the same node). This way, those processes can benefit from higher bandwidth of memory compared to network interface bandwidth. In order to do this, Application Graph (AG) and Cluster Topology Graph (CTG) are established and then, it is tried to find an efficient mapping from AG to CTG. In AG, vertices represent parallel processes and edges represent communications between processes. In CTG, vertices and edges represent processing cores and available communication bandwidth between them, respectively. Since graph mapping problem is an NP problem, some heuristics have been introduced which are based on graph partitioning approaches. Dual Recursive Bipartitioning (DRB) and K-way graph partitioning are two common heuristics. In DRB, AG is divided into two subgroups such that processes which frequently communicate to each other will be grouped in the same subgroup, but processes which communicate to each other infrequently, will be placed in different subgroups. By 'frequently' we mean the total volume of data exchanged between each pair of processes. The CTG is also divided into two subgroups in the same way as done in AG. In the next step, each subgroup of AG is assigned to the peer subgroup of CTG. This procedure is repeated on each subgroup recursively until one process in AG or one processing core in CTG remains. K-way graph partitioning is the same as DRB except that instead of two subgroups, graphs are divided into K subgroups.

Although graph partitioning algorithms offer communication performance by mapping frequently communicating processes near each other, but when we try to put such processes near themselves, some shared resources can become performance bottlenecks and these methods are oblivious to this issue. Currently, there are limited studies that propose a mapping strategy to mitigate contention problem. Among them we can point to [16-18]. [16] Introduces a mapping algorithm to avoid congestion on Torus interconnection networks. But this study does not pay to the problem of congestion on the network interface. In [17] the problem of contention on network interface is investigated. This study attempts to put a combination of parallel jobs which have high inter-node communications and low inter-node communications in one computing node. This way, network interface contention is alleviated while maximum number of processing cores is used in an efficient way. However this study does not provide a systematic algorithm to use in all scenarios and under every condition. Proposed method in [18] is based on a scheduling method to mitigate contention and does not benefit from an intelligent mapping technique.

## 4. PROPOSED MAPPING STRATEGY

For the sake of reducing contention on a network interface, the conditions in which contention is raised, must be recognized. By determining such conditions, we can present the solution. If we could accommodate all processes of a parallel job in just one computing node, there will be no usage of network interface and hence, there is no race to send inter-node messages. But when the number of parallel processes is high, or the number of available free cores in the computing nodes is low, parallel processes must be placed in more than one computing node, inevitably. In this case, high volume of inter-process communications can raise the contention on the network interface and hence, it can substantially degrade the performance of inter-node communications. To confront this problem, we should fix a threshold on the number of parallel processes which reside in a node and have high volume of inter-node communications. This means that we should distribute processes among available nodes so as to reduce network requests arriving to each interface. Consequently, waiting time at interface queue for inter-node messages will be decreased. In this study, we try to determine an appropriate value for the threshold using the number of adjacent processes of each process and the number of available free cores in the computing nodes. Figure 1 shows our proposed mapping strategy pseudocode. Although the heart of the proposed algorithm is to determine a threshold value for the number of processes in each computing node, but there are also other important steps in the mapping procedure. The first step is to separate parallel jobs based on the length of messages they send. Since larger messages require more service time, processes which send larger messages should be sent as intra-node messages to benefit from high bandwidth of memory. This can hugely reduce the waiting time of large messages which constitute a major part of communications. In doing so, we categorize messages into 3 groups: large messages (1MB or higher), medium messages (2KB to 1MB), and small messages (2KB or less). According to these categories we separate parallel jobs and because of the mentioned issue, those jobs which send large messages are first selected for mapping (step 1). After that, it is the time to select and map jobs which send medium and small messages respectively (steps 4, 6). But parallel processes of a job may send messages with different sizes. In such cases largest message length is considered for action. After partitioning jobs, parallel jobs in each group are sorted (step 2) based on average number of adjacent processes of each process ($Adj_{avg}$) and jobs which have more average adjacency are mapped earlier. This is because these jobs may need to distribute between the nodes to obtain efficient performance. As a result, these jobs should be mapped before other jobs to use available free cores of computing nodes. After choosing a job to map, processes of this job are sorted based on their total communication demands (step 3.3) and processes which have more communication demands, are mapped earlier. In the proposed strategy, communication demands for process 'i' ($CD_i$) is calculated as follows:

$$CD_i = \sum_{j=1, j\neq i}^{P} L_{ij}\lambda_{ij} \qquad (1)$$

In the equation above, $L_{ij}$ is the size of messages sent from process 'i' to process 'j' (largest length when having different lengths), $\lambda_{ij}$ is the rate of sending messages from process 'i' to 'j', and P represents the number of parallel processes for current job. After determining process with the most communication demands (given process 'A'), this process is assigned to a node with most free processing cores (steps 3.5 to 3.7). In the next step, adjacent processes of 'A' are sorted based on the communication demands between 'A' and them, and it is tried to map adjacent processes of 'A' in the same node as 'A' (step 3.9). Now, it must be noted that if the number of adjacent processes is high, or the number of available free cores in current node is low, some adjacent processes must be mapped to other computing nodes. In such situations, as mentioned earlier, high volume of inter-process communications can lead to severe contention on the network interface and degrade the performance. So before mapping processes of current job, we should determine a threshold on the number of processes which reside in a node and use network interface for their inter-node communications.

```
New_Mapping_Strategy( )
Input: Workload information, Cluster architecture
Output: Mapping information
{
 1. job_pool = select_jobs ( high_length );
 2. sort_jobs ( job_pool );
```

```
3. while ( job_pool is not empty )
{
  3.1. crnt_job = select_job ( job_pool );
  3.2. If ( Adj_avg <= FreeCores_avg -1 )
          No threshold is determined;
        Else
```

$$Threshold = \left\lceil \frac{\sum_{i=1}^{P} \frac{Adj_{pi}}{Adj_{\max}}}{num\_of\_nodes} \right\rceil ;$$

```
  3.3. sort_process ( crnt_job );
  3.4. crnt_process = select_process ( crnt_job );
  3.5. crnt_node = selec_node ( cluster_arch );
  3.6. crnt_socket = select_socket ( cluster_arch );
  3.7. map_process ( crnt_process, crnt_node, crnt_socket);
  3.8. sort_adj ( crnt_process );
  3.9. map_adj_processes ( threshold );
}
4. job_pool = select_jobs ( medium_length );
5. repeat steps 2,3;
6. job_pool = select_jobs ( small_length );
7. repeat steps 2,3;
}
```

Figure 1. Pseudocode of the proposed mapping strategy

To determine the threshold, we act as follows: If average adjacency for parallel processes is less than or equal to the average number of free processing cores (FreeCores$_{avg}$) in available computing nodes (except one processing core which is used to map process 'A' to it), we can say roughly that 'A' and its adjacent processes can reside in just one node and there is no significant inter-node communications, probably. In this case, there is no need to fix a threshold value. In contrast, if average adjacency is higher than the average free cores, some processes must be placed out of the current node. In this case, threshold is determined by:

$$Threshold = \left\lceil \frac{\sum_{i=1}^{P} \frac{Adj_{pi}}{Adj_{\max}}}{num\_of\_nodes} \right\rceil \qquad (2)$$

In equation 2, a weight ($\frac{Adj_{pi}}{Adj_{\max}}$) is assigned to each process. In this weighted value, Adj$_{pi}$ represents number of adjacent processes for process pi and Adj$_{max}$ represents maximum adjacency between parallel processes within current job. The reason for choosing a weighted threshold value is because high amount of adjacency makes us determine a threshold. Consequently, processes which have more adjacency should have more contributions (or weight, as a result) on selected threshold value than others. The sum of weighted values is then divided by the number of computing nodes (num_of_nodes) to distribute processes between all computing nodes. It is to be mentioned that although distributing processes between all cluster nodes, does not always lead to optimum results, but our experiments revealed that in many scenarios, it would yield efficient performance. An important note about equation 2 is that if the number of computing nodes is more than the number of parallel processes, the threshold will be equal to 0 which is meaningless. In this case, we set the threshold value to 1.

International Journal of Grid Computing & Applications (IJGCA) Vol.3, No.2, June 2012## 5. EVALUATION OF THE NEW MAPPING STRATEGY

### 5.1. Simulation Testbed

To perform our experiments, we used Omnet++ v4.1 simulator. The simulated platform is a multi-core cluster containing 16 computing nodes which are connected together through an intermediate switch. Each computing node has 4 sockets and each socket is a quad-core processor, so each node contains 16 processing cores. The architecture of each node is based on the NUMA[1] architecture. This means that each socket can access to its local memory (although it can also access to remote memories but with more latency). In each node, we used a network interface with InfiniBand technology. InfiniBand is one of the most advanced interconnection networks which is used to establish high performance clusters. Table 1 lists the parameters we used in our simulations.

Table 1. Simulation Parameters

| Parameter | Value |
| --- | --- |
| Main memory bandwidth | 4GB/s |
| Remote memory access latency | 10% more than local memory access latency |
| Cache bandwidth (for intra-chip communications) * | Corresponds to AMD Opteron 2352 chip |
| Maximum size of common buffer in cache ** | 1MB |
| Network interface bandwidth | 1GB/s (corresponds to InfiniHost MT23108 4x) |
| Switching latency at intermediate switch | 100ns (independent of message size) |

* Processing cores that reside in the same socket, can benefit from intra-chip cache for their message communications.

** Because of the limited capacity of the cache, maximum transferable message size using the cache is 1MB, and larger messages should be transferred through the main memory.

### 5.2. Experimental Results for Synthetic Workloads

In order to evaluate our new mapping strategy, we used two sets of workloads: synthetic workloads and real workloads. In this section we show the performance results for synthetic workloads and the results for real workloads are shown in the next section. In synthetic workloads, messages which had different lengths and rates were generated. In these traffics, we defined four different communication patterns between parallel processes. These patterns which are common communication patterns in message passing libraries are: Gather/Reduce, Bcast/Scatter, Linear and All-to-All. In Gather/Reduce pattern, one process as the root process, receives messages from other processes and other processes are just senders. In Bcast/Scatter pattern, one process as the root process sends its messages to other processes and other processes are just receiver. In Linear pattern, each process receives messages from a previous process and sends its messages to a next process (there is a linear communication pattern between processes). Finally, in All-to-All pattern, each process sends messages to all other processes. Tables 2 to 5 show the definition of 4 synthetic workloads which each, contains a number of parallel jobs with different communication patterns.

---

[1] Non Uniform Memory Access

34

Table 2. Synt_workload_1

| Job | No. of Processes | Pattern | Length | Rate | Message Count |
|-----|------------------|---------------|--------|--------|---------------|
| 0 | 64 | All-to-All | 64KB | 100m/s | 2000 |
| 1 | 64 | Bcast/Scatter | 64KB | 100m/s | 2000 |
| 2 | 64 | Gather/Reduce | 64KB | 100m/s | 2000 |
| 3 | 64 | Linear | 64KB | 100m/s | 2000 |

Synt_workload_1 is interpreted as follows: This workload has 4 parallel jobs and each job has 64 processes. Each process sends messages with specific characteristics (in terms of length, rate and communication pattern). Each process finishes its execution after sending a specific number of messages (Message Count). Other synthetic workloads have similar interpretation.

Table 3. Synt_workload_2

| Job | No. of Processes | Pattern | Length | Rate | Message Count |
|-----|------------------|---------------|--------|-------|---------------|
| 0 | 64 | All-to-All | 2MB | 10m/s | 2000 |
| 1 | 64 | Bcast/Scatter | 2MB | 10m/s | 2000 |
| 2 | 64 | Gather/Reduce | 2MB | 10m/s | 2000 |
| 3 | 64 | Linear | 2MB | 10m/s | 2000 |

Table 4. Synt_workload_3

| Job | No. of Processes | Pattern | Length | Rate | Message Count |
|-----|------------------|---------------|--------|-------|---------------|
| 0 | 32 | All-to-All | 2MB | 10m/s | 2000 |
| 1 | 32 | Bcast/Scatter | 2MB | 10m/s | 2000 |
| 2 | 32 | Gather/Reduce | 2MB | 10m/s | 2000 |
| 3 | 32 | Linear | 2MB | 10m/s | 2000 |
| 4 | 32 | All-to-All | 64KB | 10m/s | 2000 |
| 5 | 32 | Bcast/Scatter | 64KB | 10m/s | 2000 |
| 6 | 32 | Gather/Reduce | 64KB | 10m/s | 2000 |
| 7 | 32 | Linear | 64KB | 10m/s | 2000 |

Table 5. Synt_workload_4

| Job | No. of Processes | Pattern | Length | Rate | Message Count |
|-----|------------------|---------------|--------|-------|---------------|
| 0 | 24 | All-to-All | 2MB | 10m/s | 2000 |
| 1 | 24 | Bcast/Scatter | 2MB | 10m/s | 2000 |
| 2 | 24 | Gather/Reduce | 2MB | 10m/s | 2000 |
| 3 | 24 | Linear | 2MB | 10m/s | 2000 |
| 4 | 24 | All-to-All | 64KB | 10m/s | 2000 |
| 5 | 24 | Bcast/Scatter | 64KB | 10m/s | 2000 |
| 6 | 24 | Gather/Reduce | 64KB | 10m/s | 2000 |
| 7 | 24 | Linear | 64KB | 10m/s | 2000 |

For performance evaluation, we used sum of the waiting times of messages at server queues (network interface and memory) as our main metric. We compared our results with the results obtained from Blocked, Cyclic and DRB methods. Figure 2 shows the performance results for 4 synthetic workloads. In this figure, 'B' indicates Blocked, 'C' indicates Cyclic, 'D' indicates

DRB, and 'N' indicates our new mapping method. To extract the results for DRB method, we used Scotch v5.1 software which implements this method

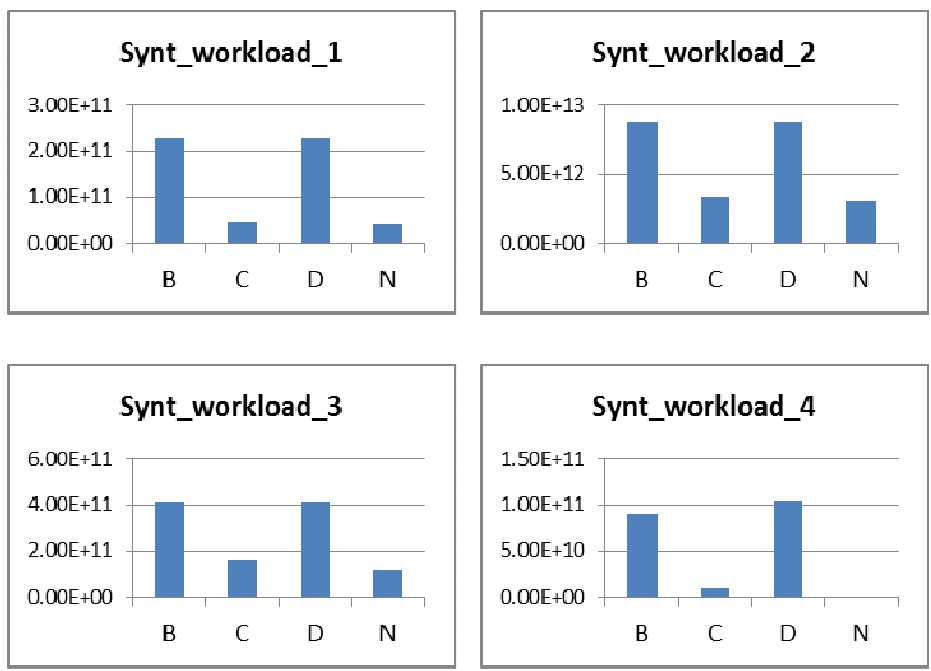

Figure 2. Waiting time of messages for synthetic workloads (in mili-seconds)

According to figure 2 we can see that the new mapping strategy has produced better results compared to the other methods. In synthetic workloads, there are jobs with All-to-All pattern which is a communication-intensive pattern. Moreover, the number of processes in each parallel job is more than the number of processing cores within a node (16 cores). These factors cause synthetic workloads to be heavy communicating workloads. In such workloads, the Blocked technique which tries to accommodate parallel processes in the minimum number of nodes, has led to severe contention on the network interface and has inefficient performance. In contrast, the Cyclic has gained better performance by distributing processes among the computing nodes. Since in the DRB method, parallel processes which are communicating frequently, are mapped near each other, process mapping is done as Blocked and the results are not efficient. The reason that the new method has performed more efficient than the Cyclic is that in the new algorithm, efficient mapping conditions is determined for each parallel job independent of the other jobs. In other words, if there are high amount of adjacency and communications between processes, the new method will distribute parallel processes among computing nodes, otherwise it acts like Blocked. Based on the performance results, the new mapping technique has gained performance improvements up to 5%, 8%, 29% and 91% for Synt_workload_1 to Synt_workload_4, respectively (performance gain is calculated compared to the best result from the other methods, i.e. Cyclic in here).

Although we used waiting time of messages at server queues as our main metric, but we can also employ other important metrics for performance comparisons. Two other important metrics are: workload finish time (the time at which execution of all parallel jobs in the workload is finished) and total finish time of parallel jobs (the sum of finish times for all jobs in a given workload). Performance results using these two metrics are illustrated in figures 3 and 4 (the results are in second (sec) unit). Once again, we can see that the new mapping method has generated better results in almost all scenarios, compared to the other methods.

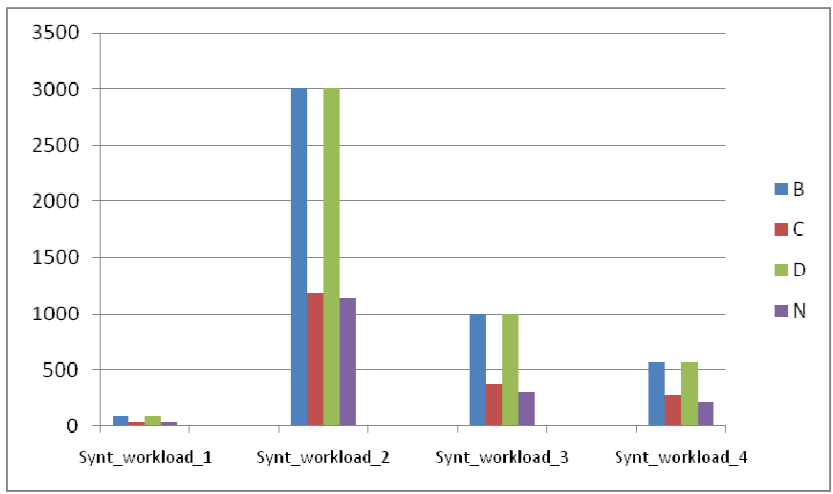

Figure 3. Workload finish time for synthetic workloads

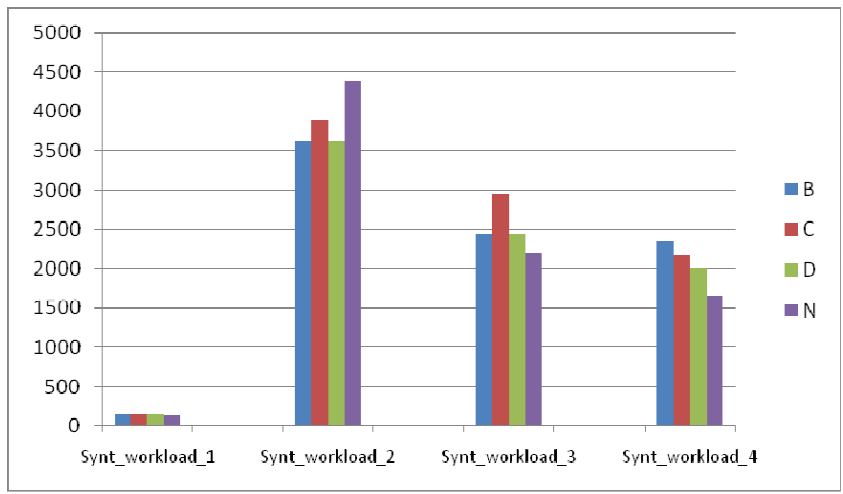

Figure 4. Total finish time of parallel jobs for synthetic workloads

## 5.3. Experimental Results for Real Workloads

Real workloads were extracted from communication behaviour of NPB[2] benchmarks. Tables 6 to 9 show the definition of 4 real workloads which each, contains some benchmarks with different number of processes and different benchmark classes. The performance results for real workloads are shown in figure 5.

---

[2] NAS Parallel Benchmarks

International Journal of Grid Computing & Applications (IJGCA) Vol.3, No.2, June 2012

Table 6. Real_workload_1

| Job | No. of Processes | Benchmark | Class |
|---|---|---|---|
| 0 | 25 | SP | C |
| 1 | 32 | IS | C |
| 2 | 32 | FT | B |
| 3 | 16 | FT | B |
| 4 | 16 | IS | C |
| 5 | 32 | CG | C |
| 6 | 8 | IS | B |
| 7 | 25 | BT | C |
| 8 | 16 | CG | B |

Table 7. Real_workload_2

| Job | No. of Processes | Benchmark | Class |
|---|---|---|---|
| 0 | 8 | IS | B |
| 1 | 32 | FT | B |
| 2 | 32 | IS | C |
| 3 | 32 | MG | C |
| 4 | 32 | CG | C |
| 5 | 32 | IS | B |
| 6 | 32 | MG | B |
| 7 | 32 | CG | B |
| 8 | 16 | BT | C |

Table 8. Real_workload_3

| Job | No. of Processes | Benchmark | Class |
|---|---|---|---|
| 0 | 25 | BT | B |
| 1 | 32 | CG | B |
| 2 | 32 | EP | B |
| 3 | 32 | FT | B |
| 4 | 32 | IS | B |
| 5 | 25 | LU | B |
| 6 | 32 | MG | B |
| 7 | 25 | SP | B |

Table 9. Real_workload_4

| Job | No. of Processes | Benchmark | Class |
|---|---|---|---|
| 0 | 25 | SP | C |
| 1 | 32 | CG | C |
| 2 | 32 | EP | C |
| 3 | 32 | MG | C |

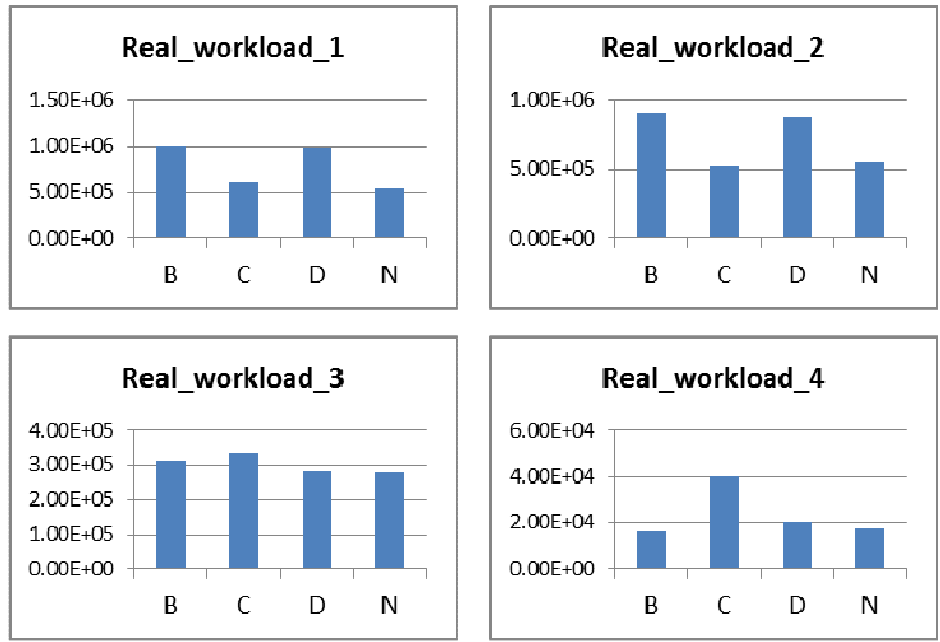

Figure 5. Waiting time of messages for real workloads (in mili-seconds)

Real_workload_1 and Real_workload_2 scenarios are communication-intensive benchmarks since they use IS and FT benchmarks more than other benchmarks. These benchmarks have high communication volume and their communication pattern is entirely in the form of All-to-All pattern. Consequently, the above mentioned workloads are heavy workloads in terms of communications. As can be seen from figure 5, the Cyclic method has performed better than the Blocked and DRB methods. In these workloads, the new approach has acted as efficient as the Cyclic and even better (in Real_workload_1 scenario, 11% performance improvement is observed).

In order to show that our strategy can produce efficient results not only in heavy workloads, but also in non-heavy workloads, Real_workload_3 and Real_workload_4 were defined. Real_workload_3 is a medium workload in terms of communications and as can be seen in figure 5, there is no significant difference between performance results of different methods for this scenario. Despite this, the new algorithm has performed a little bit better than the others. Real_workload_4 is a scenario which has light communication demand and as we can expect, the Blocked and DRB methods have achieved better results compared to the Cyclic. Performance results for this scenario show that the new mapping method has performed as well as Blocked which indicates that the new approach can have efficient results even in light communicating workloads.

## 6. CONCLUSION

In this paper, we proposed a new process mapping strategy to use in multi-core cluster environments. The goal of the proposed technique is to assign parallel processes to processing cores aimed at reducing network interface contention. Since the number of processing cores within recent computing nodes is growing up with a rapid rate, contention on shared resources is posing itself as a serious challenge and should be tackled for enhancing the performance. Here, we confronted this problem and proposed a process placement method to alleviate contention on network interfaces as one of the main shared resources. We compared our technique with some other well-known methods and significant improvement in performance was gained (5% to 90%) in experimental workloads. Our mapping method is easy to implement and its efficiency makes it usable in recent high performance multi-core cluster environments.

## 7. FUTURE WORKS

Many aspects of this study can be extended as future work. The first one is to consider computation demands of parallel processes as well as their communication demands. In the proposed technique we assumed that the cluster environment is a homogenous system in which all computing nodes are equal in terms of architecture and their computational capabilities. But real environments may use different nodes with different computational speeds. In such cases, we should not only consider the communication demands of parallel processes, but we should also take into account the computational efficiency of parallel jobs and computation-intensive processes should be mapped to faster computing nodes. This issue can be considered to extend the proposed method as a future work.

Another aspect to extend the work is to change the scenario used for the clustered system. The system which we considered here, was a high performance cluster (HPC) aimed at running parallel applications with high performance. But we can consider other scenarios for the cluster such as a high throughput scenario and investigate the problem in this new case. In a high throughput cluster such as a web server, different processing cores do not necessarily communicate to each other, but all of them may need to deliver the results using the network interface and the contention problem on the interface may be more challenging than before.

## ACKNOWLEDGEMENTS

We thankfully appreciate Research Institute for ICT (ITRC) of Iran for their support. We hope this work to be a valuable research for extending the technological knowledge of this institute.

**Authors**

**Mohsen Soryani** received his B.Sc. in Electronics Engineering from Iran University of Science and Technology, Tehran, Iran in 1980. He also received his M. Sc. and Ph.D. in Digital Techniques and Image Processing respectively from Heriot-Watt University, Edinburgh, Scotland in 1986 and 1990. He is currently an assistant professor in the School of Computer Engineering, Iran University of Science and Technology. His research interests include computer architecture and networking, networks on chip and image and video processing.

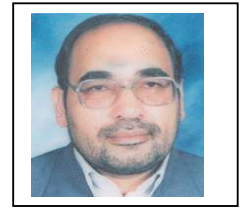

**Morteza AnaLoui** received a B.A. degree in Electrical Engineering from Iran University of Science and Technology (IUST) in 1983 and a Ph.D. degree in Electrical Engineering from Okayama University, Japan, in 1989. He is an Associate University Professor (and past Dean) in the College of Computer Engineering and the director of Computer Network Research Lab in IUST. Dr. AnaLoui has been an assistant professor in the Faculty of Engineering, Tarbiat Modaress University, 1994 and in the Engineering Faculty of Okayama University from 1990 to 1993. His research interests include Networking (High Speed, Virtual Social, Virtual Economic, CDN, P2P), Optimization Theories and Techniques, Pattern and Cognitive recognition, Trust management and Persian Computing. Dr. AnaLoui has authored over 150 publications, including journal articles, book chapters, and conference papers.

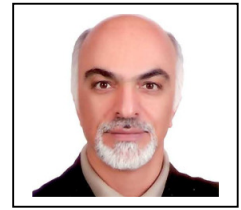

**Ghobad Zarrinchian** received his B.E. degree in Computer Hardware Engineering from Islamic Azad University (Mashhad branch) in 2008. He also received his M.Sc. degree in Computer Systems Architecture Engineering from Iran University of Science and Technology (IUST) in 2011. His research interests include computer architecture, parallel processing and distributed systems.

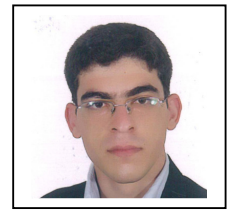